\documentclass{iau_FM}
\usepackage{graphicx}

\title[JD 11.~~The Galactic magnetic field and cosmic ray lensing] 
{The Galactic magnetic field and its lensing of ultrahigh energy and Galactic cosmic rays}

\author[Glennys R. Farrar]   
{Glennys R. Farrar
}

\affiliation{Center for Cosmology and Particle Physics, Department of Physics \\New York University, NY, NY USA}

\pubyear{2015}
\jname{Astronomy in Focus, Volume 1} 
\editors{Piero Benvenuti, ed.}
\begin{document}

\maketitle

\begin{abstract}
It has long been recognized that magnetic fields play an important role in many astrophysical
environments, yet the strength and structure of magnetic fields beyond our solar system have been at best only qualitatively constrained.   The Galactic magnetic field in particular is crucial for modeling the transport of Galactic CRs, for calculating the background to dark matter and CMB-cosmology studies, and for determining the sources of UHECRs.   This report gives a brief overview of recent major advances in our understanding of the Galactic magnetic field (GMF) and its lensing of Galactic and ultrahigh energy cosmic rays.  

\end{abstract}
\keywords{ISM: magnetic fields, cosmic rays, Galaxy: halo, diffusion}

\firstsection 
\section{The Galactic Magnetic Field}

Early qualitative indications of large scale structure in galactic magnetic fields came from external galaxies and the pattern of Rotation Measures of pulsars;  today we have a quantitative description of the magnetic field of our Galaxy, including coherent, random and ``striated-random'' components.  The current leader in field models in terms of the generality of the functional form and its ability to reproduce the constraining observational data, is the model of Jansson \& Farrar (2012a,b) (JF12).   Parameters of the model were determined by fitting all-sky Faraday Rotation Measures of $\approx$40k extragalactic sources, simultaneously with WMAP polarized (Q,U) and total synchrotron emission maps;  this data together provides a total of more than 10,000 independent datapoints, each with measured (predominantly, astrophysical) variance.

\begin{figure}[b!]
\centering
\includegraphics [width=0.5\columnwidth, angle=-90]{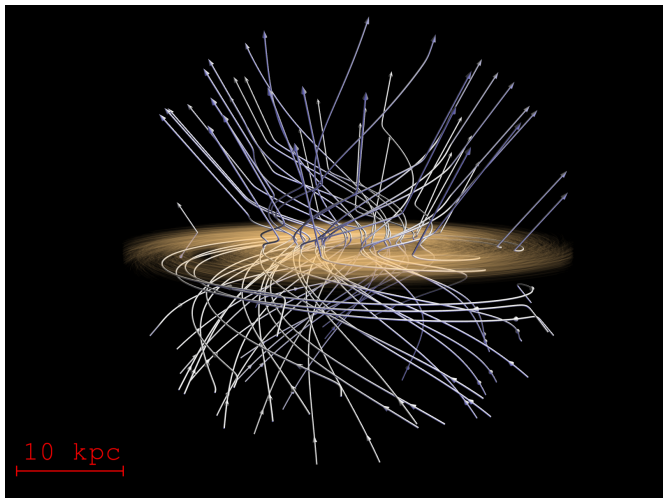}
\caption{The outwardly-spiraling halo field of the Galaxy (JF12); the field lines run from S to N.   Visualization produced by T. Sandstrom, NASA. }
\label{GMF}
\end{figure}


In addition to disk and toroidal halo components, the JF12 field model allows for a coherent poloidal halo component, which proved essential for fitting the data.  In addition, JF12 allowed for a ``striated" random component, defined to be a component which averages to zero along a typical line of sight, but which has a net  preferred direction such that it contributes to polarized synchrotron emission.   An example would be the field produced by stretching or compressing a random field, e.g., due to a wind, or by a supernova explosion within a region of coherent field.   Jansson and Farrar (2012a) find their best fit when the orientation direction of the striated field is locally aligned with the direction of the coherent field, so that the striated contribution amounts simply to an enhancement of the polarized synchrotron signal relative to the RM signal.  This suggests that compression of a coherent field, e.g., by SNe explosions, may be the origin of the striated field effect (Farrar, 2014).   The full JF12 model depends on 34 parameters: 20 for the coherent field, one for the striated field, and 13 for the random field (Jansson and Farrar, 2012b) and gives an excellent fit to the observables.  A more detailed discussion of the JF12 model and comparison to data and other models can be found in \cite{fCRAS14};  for further results on variants of the JF12 field and on the robustness of the JF12 description see \cite[Khurana \& Farrar (2015)]{kf15}.

The observational data on RM, Q, U unambiguously demand a large scale pinched-spiral halo field with field lines running from South to North.  The relative signs of the poloidal and toroidal components are consistent with what would be produced by differential shear in the Galaxy, starting with the N-to-S poloidal component.  Fig. \ref{GMF} shows the field lines of the coherent halo field in the JF12 model; the field of the disk is rendered with finer but more dense lines in a different color and without directional arrows, to aid visual clarity.   \vspace{-1pc}

\begin{figure}
\vspace{-2.8in}
\centering
\includegraphics[width=0.9\textwidth]{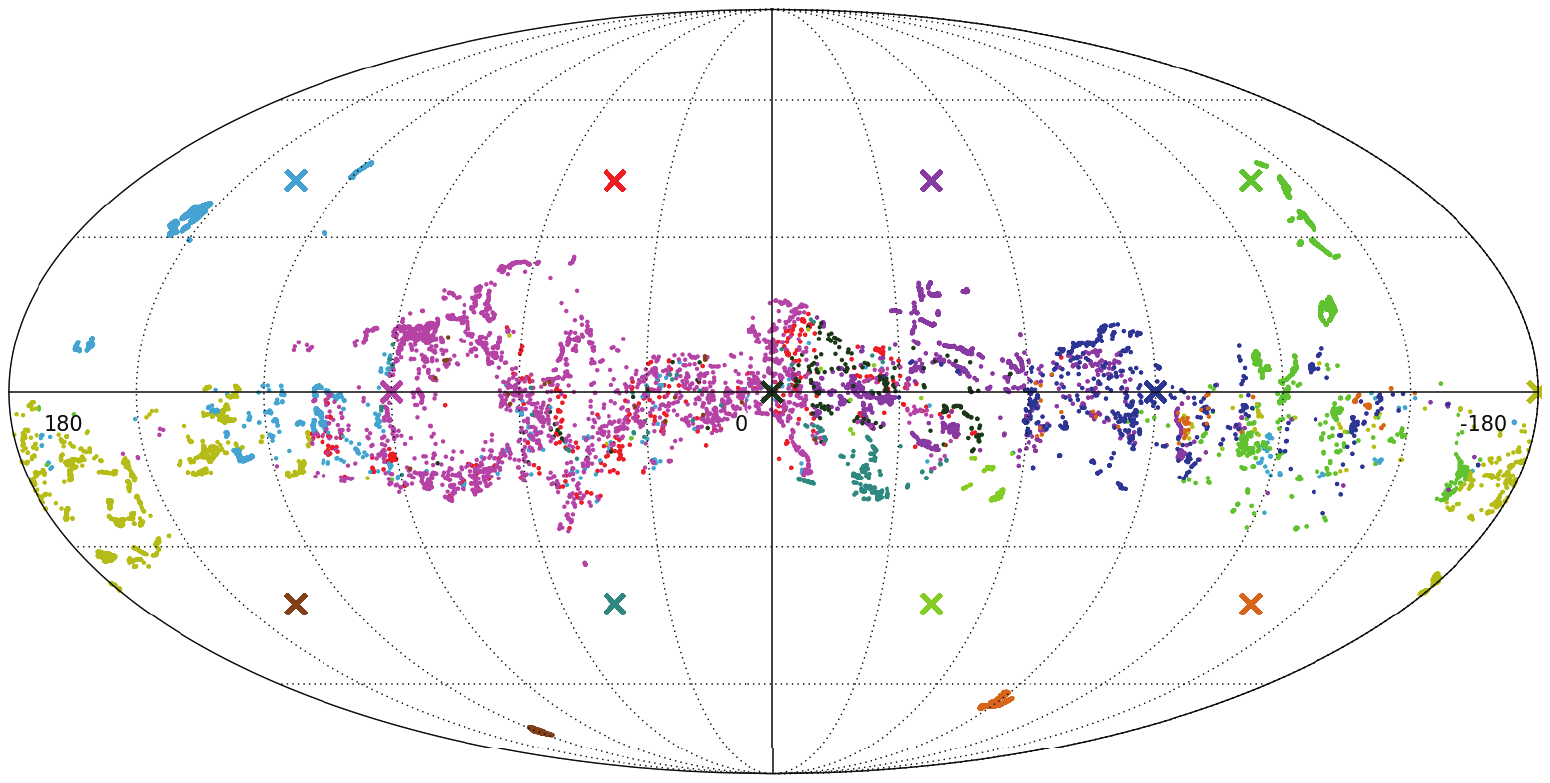}
\vspace{-2.4in}
\caption{Arrival directions of 10 EV CRs, for 12 representative source directions (marked with crosses) at b = $\pm 41.5^{\circ}$, l = $\pm 45^{\circ}$ and $\pm 135^{\circ}$, and at b=0, l = 0, $\pm 90^{\circ}$ and $180^{\circ}$; events from a given source have the same color as the source marker.}  
\label{magdefs}
\end{figure}

\section{CR deflections in the GMF}


UHECR lensing produces multiple images and dramatic magnification and
demagnification that varies with source direction and CR rigidity, E/Z.  Galactic CR propagation is significantly anisotropic in the JF12 field, because the halo field provides a heretofore-not-considered escape route via diffusion along its field lines.   
\smallskip

\noindent {\underline{\bf  UHECR deflections:}}
\\Fig.  \ref{magdefs} shows the observed arrival directions for 12 selected source directions (the centers of the 12 base-level Healpix pixels)  at E/Z = 10$^{19}$V.  One sees that deflections are generally very large.  Except for the 4 directions farthest from the GC (b = $\pm 41.5^{\circ}$, l = $ \pm 135^{\circ}$), the arrival directions are typically highly dispersed.  The Galactic plane is something of an ``attractor'', apart from the directions in the south and away from the galactic center, for which the deflections cause the arrival directions to come from further to the S. By comparing different realizations of the random field, one sees -- not surprisingly -- that the average deflection is basically the same, as it depends mainly on the coherent field, while the degree of dispersion decreases when the coherence length or $B_{\rm rand}$ is reduced.  Space does not permit inclusion of more examples, but as would be expected, mean deflections generally increase with decreasing rigidity.  At higher rigidity, such that the deflections do not carry the UHECR too close to the  Galactic plane, the deflections can be much smaller, especially for directions away from the Galactic center.  UHECR deflection maps for arbitrary source directions and rigidities from $10^{18}-10^{20}$V, and movies showing the trajectories themselves, are available;  contact GRF.

The GMF acts as a lens for UHECRs (Harari {\em et al.}, 2000).  Although the total number of UHECRs is conserved, and an isotropic distribution of sources leads to an isotropic observed sky, the total flux from any given source direction can be magnified or demagnified relative to the flux in the absence of the GMF, because smaller or larger areas of the plane wave of UHECRs coming from a given source direction can be focused onto the Earth;  for illustrative figures see \cite{fCRAS14}.  In the case that multiple distinct regions of the source plane are focussed onto Earth, multiple images of the source are produced, as can be seen in Fig. \ref{magdefs} and discussed in \cite{kfs14} for Cen A as the source.  

\begin{figure}[!]
\centering
\includegraphics [width=0.85\columnwidth]{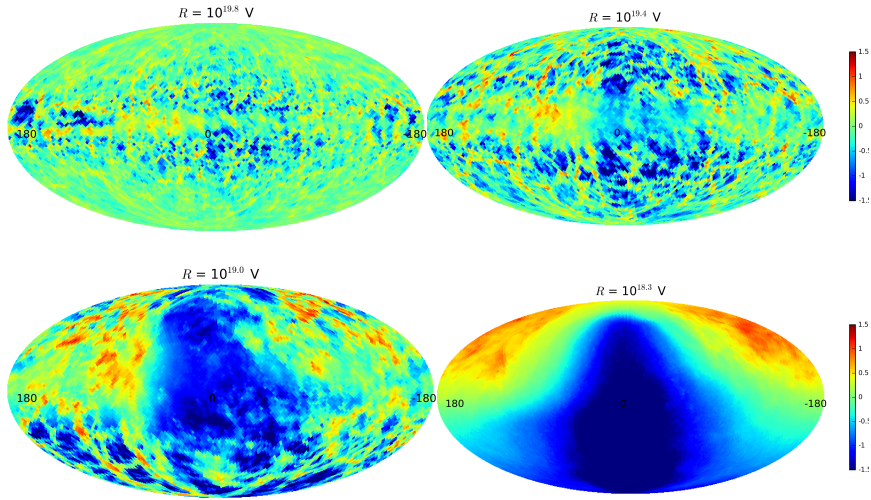}
\vspace{-4pc}
\caption{Skymaps showing log$_{10}$ of the magnification, for 4 different rigidities.  Blue means no CRs from that source direction reach Earth (they are deflected away from us by the GMF) while a source direction indicated with the deepest red has a factor 30 magnification.  For some source directions there is a rapid shift from magnification to demagnification as the rigidity changes.}
\label{magmap}
\end{figure}

While the existence of magnetic lensing is to be expected, the actual impact of the Milky Way magnetic field on UHECR arrival directions can only now be explored, thanks to new high-resolution simulations and use of a realistic field model (Farrar \& Sutherland, 2015).  The results are quite remarkable.  Fig. \ref{magmap} shows skymaps of log$_{10}$ of the magnification at various rigidities, as a function of source direction on the sky.  Dark blue represents complete blindness to the given direction -- UHECRs from those directions are deflected away by the GMF and totally miss Earth -- while the darkest red represents a magnification by a factor 30.   
As the random field is changed, e.g., reduced in strength or the coherence length changed, the details of the pattern of magnification changes, but qualitatively the picture is the same.  The most important features of the magnification map are: 
\begin{itemize}
\item  At high rigidity, the angular size of high-magnification and blind regions is small and their position varies with random field model.
\item For a given random field configuration, the magnification and demagnification for a given source direction can vary rapidly with rigidity.
\item As the rigidity decreases, the blind region grows until about half the sky is invisible --  from behind the Galactic center and especially from the South.
\end{itemize}  
\smallskip

\begin{figure}[t!]
\centering
\vspace{-1.5 pc}
\includegraphics [width=0.7\columnwidth, angle=-90]{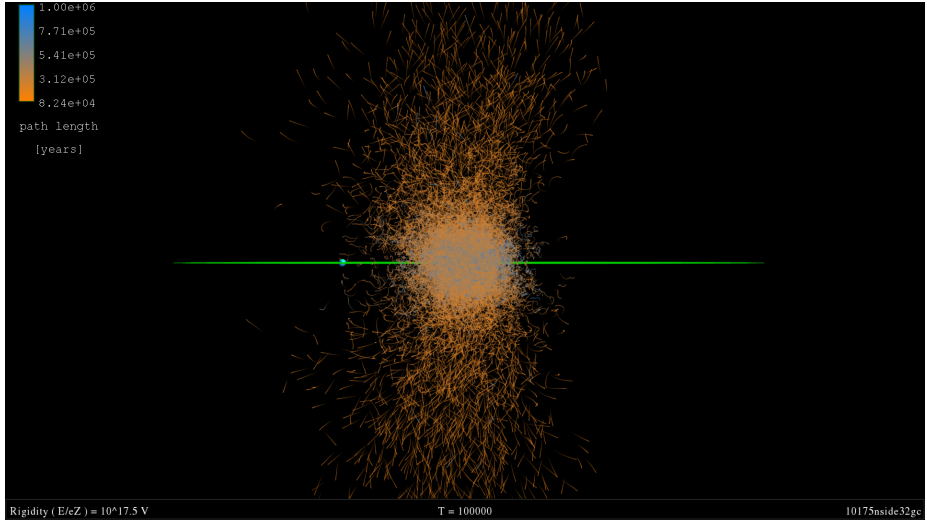}
\vspace{-2.5pc}
\caption{The tracks of high energy (rigidity = E/Z = 0.3 EV) Galactic Cosmic Rays produced at the Galactic Center.  Note the strong anisotropy in diffusion -- almost no cosmic rays reach Earth before escaping the Galaxy.  
}
\label{GCsource}
\end{figure}

\noindent {\underline{\bf  Galactic CR diffusion:}}  \\
Fig. \ref{GCsource} shows a snapshot of 0.3 EV CRs tracked through the GMF from an explosion at the Galactic Center, 100 kyr after the explosion.  The important point to notice is how anisotropic the ``diffusion'' is -- preferentially escaping vertically.  Almost no CRs reach the solar circle at these energies.   
While our Galactic CR simulations were done for very high rigidities, the propagation pattern is similar for CRs down to much lower energies relevant for Fermi diffuse emission and dark matter annihilation searches.  This is because the propagation we observed is quasi-diffusive and to a good approximation the rigidity dependence of the diffusion tensor factorizes from its tensor structure, the latter depending on the local random and coherent fields.  Thus these simulations capture the spatially varying anisotropy of the diffusion tensor which is not property treated in standard codes such as GALPROP and DRAGON.  For lower rigidities, the time it takes to reach a given distance from a source would increase, but not the spatial distribution.  These results underline the crucial importance of including anisotropic diffusion in a realistic magnetic field, before drawing conclusions about galactic cosmic rays, at any energy.  

Galactic CR trajectories have been calculated for sources at a variety of locations and for a range of CR rigidities; movies showing the time development for a transient source are available;  contact GRF.

\section*{Acknowledgments}
This research was supported in part by the U.S. National Science Foundation, NSF-PHY-1212538, and the James Simons Foundation;  special thanks to N. Awal, D. Khurana, and M. Sutherland for their contributions.


\begin{thebibliography}{1}

\bibitem[Jansson \& Farrar (2012a)]{jf12a}
R.~{Jansson} and G.~R. {Farrar}.
\textit{ApJ}, 757:14, 2012.

\bibitem[Jansson \& Farrar (2012b)]{jf12b}
R.~{Jansson} and G.~R. {Farrar}.
\textit{ Astrophys. J.}, 761:L11, 2012.

\bibitem[Khurana \&Farrar (2015)]{kf15}
D.~{Khurana} and G.~R. {Farrar}.
\newblock {In preparation}, 2015.

\bibitem[Farrar (2014)]{fCRAS14}
G.~R. {Farrar}.
\textit{Comptes Rendus Physique}, 15:339--348, April 2014.

\bibitem[Harari \etal\ (2000)]{hmrLensing00}
D.~{Harari}, S.~{Mollerach}, and E.~{Roulet}.
\textit{Journal of High Energy Physics}, 2:35, February 2000.

\bibitem[Keviani \etal\ (2014)]{kfs14}
A.~{Keivani}, G.~R. {Farrar}, and M.~{Sutherland}.
\textit{Astroparticle Physics}, 61:47--55, February 2015.

\bibitem[Farrar \& Sutherland (2015)]{fs15}
G.~R. {Farrar} and M.~{Sutherland}.
\newblock {In preparation}, 2015.

\end{thebibliography}
\end{document}